\documentclass{article} 
\usepackage{url,hyperref,microtype,subcaption}
\usepackage{graphicx, natbib, a4, amsmath, amssymb, fancyhdr, setspace, subcaption,authblk,cite,makecell}
\usepackage[a4paper, total={6in, 8in}]{geometry}
\usepackage[dvipsnames]{xcolor}
\usepackage{lscape}

\begin{document}
\onecolumn
%\firstpage{1}

\title{Effect of stress on cardiorespiratory synchronization of Ironmen athletes}

\author[$^{1}$]{Maia Angelova \footnote{Corresponding author. e-mail: maia.a@deakin.edu.au}}
\author[$^{2}$]{Philip M. Holloway}
\author[$^{1}$]{Sergiy Shelyag}
\author[$^{1}$]{Sutharshan Rajasegarar}
\author[$^{3}$]{H.G. Laurie Rauch}

\affil[$^{1}$]{D2I Research Centre, School of IT, Deakin University, Geelong, Victoria, Australia}
\affil[$^{2}$]{Northumbria University, Newcastle Upon Tyne, UK}
\affil[$^{3}$]{Department of Human Biology, Faculty of Health Sciences, University of Cape Town, Cape Town, South Africa}

%%%%%%%%%%%%%%%%%%%
\maketitle
%%%%%%%%%%%%%%%%%%%

\begin{abstract}
The aim of this paper is to investigate the cardiorespiratory synchronization in athletes subjected to extreme physical stress combined with a cognitive stress tasks. ECG and respiration were measured in 14 athletes before and after the Ironmen competition. Stroop test was applied between the measurements before and after the Ironmen competition to induce cognitive stress.  Synchrogram and empirical mode decomposition analysis were used for the first time to investigate the effects of physical stress, induced by the Ironmen competition, on the phase synchronization of the cardiac and respiratory systems of Ironmen athletes before and after the  competition. 
%The cardiorespiratory interactions were investigated during 
A cognitive stress task (Stroop test) was performed both pre- and post-Ironman event in order to prevent the athletes from cognitively controlling their breathing rates. Our analysis showed that cardiorespiratory synchronization increased   post-Ironman race  compared to  pre-Ironman.
The results suggest that the amount of stress the athletes are recovering from post-competition is greater than the effects of the Stroop test.  This indicates that the recovery phase after the competition is more important for restoring and maintaining homeostasis, which could be another reason for stronger synchronization.
%Our results suggest that respiration is a key for improving the feedback between the cardiac and respiratory systems and possibly for improving the performance of the athletes. 
\end{abstract}

%%%%%%%%%%%%%%%%%
\maketitle
%%%%%%%%%%%%%%%%%%
\section{Introduction}
An Ironman race is a long-distance triathlon consisting of a 2.4-mile swim, a 112-mile bicycle ride and a 26.2-mile marathon run raced in that order with no break in between sections. It takes a participant a long time to recover from the physiological stress of completing an Ironman race. Such heavy exertion undoubtedly has a negative effect on the body’s immune system with sustained inflammatory response to muscle fatigue, increased risk of respiratory tract infections, weight loss, and other medical conditions \citep{Ren2019}. In this study we measured the effect of extreme physical stress on the cardiorespiratory system of athletes, while they performed a cognitive test to prevent them from cognitively controlling their breathing rates. 

We studied the performance of the cardiorespiratory system by investigating the effects of the Ironmen competition on cardiorespiratory synchronization. The cardiovascular and respiratory systems are coupled by several mechanisms two di\citep{Berne1998,Ren2019}, where the interactions between these two systems involve a large number of feedback and feed-forward mechanisms. In healthy subjects, the heart rate increases during inspirations and decreases with expiration, which is the well-known, and well-studied  phenomena \citep{Anrep1936} respiratory sinus arrhythmia (RSA). However, for cardiorespiratory system, it is unlikely to  find continuous synchronization, as the  respiration is neither the governing mechanism, nor is the only system affecting the heart rate dynamics. In fact, synchronizations is only expected to be observed either when one of the systems is forced (i.e., by controlled breathing) or when synchronizations are necessary for the regulation of homeostasis (i.e. after events that induce stress in one or both of the systems).
%%%%%%%%%%%%
%%%%%%%%%%%%%

Earlier studies of cardiorespiratory synchronization support its existence \citep{Pokrovskii1985, Wu2006, Bartsch2007, Rosenblum2004, Angelova2017} and as shown in \citep{Bartsch2012} cardiorespiratory synchronization and RSA represent different aspects of the interaction between the cardiac and respiratory systems. Cardiorespiratory synchronizations were shown to exist in humans during rest \citep{Schafer1998, Lotric2000, Stefanovska2001,Ren2019}, Zen meditation  \citep{Cysarz2005}, Dharma-Chan meditation \citep{Chang2013}. Desynchronizations were reported following myocardial infarctions \citep{Leder2000, Hoyer2002}, as well as reduced cardiorespiratory coordination with obstructive sleep apnoea \citep{Kabir2010} and acute insomnia \citep{Angelova2020}.

As with most physiological time series, when investigating the coupling of the cardiorespiratory systems, noise will occur.  This noise originates not only from measurements and external disturbances, but also from the fact that there are other subsystems that take part in the cardiovascular control \citep{Stefanovska1999,Angelova2020}. These influences, when considering cardiorespiratory synchronizations, are also considered as noise.

Cognitive stress is known to affect the physiological functioning of the cardiovascular system suppressing heart rate variability (HRV)  \citep{Hansen2003, Wood2002,Ren2019}.  In physiology, HRV is the variation in time interval between heartbeats, measured by the variation in the beat-to-beat interval \citep{Hon65}.   Raschke et al. suggested that coordination between the cardiac and respiratory systems would be at its strongest during states of relaxation and stated that this coordination was easily disturbed under conditions of stress or disease \citep{Raschke1987}. However, there is little knowledge of  the effect that cognitive stress exerts on cardiorespiratory synchronizations and no study thus far has investigated neither the effect of extreme physical stress on cardiorespiration, nor of cognitive stress during or after an extreme physical stress such as the Ironmen competition. In this  study, the participants were asked  to complete a Stroop test in order to impose stress and draw attention away from consciously controlling one's breathing and instead focus on completing the task.
Our hypothesis is that we will see a decrease in the amount of synchronization during the Stroop test, due to the physical stress of the Ironman event.

After the first Stroop test, the participants completed the Ironmen competition, after which a second Stroop test was administered. Thus, we  observed the effect of the extreme physical task on the concentration and cognitive abilities. Coordination between the cardiorespiratory systems has been reported in healthy adults \citep{Lotric2000,Kotani2002}, athletes \citep{Schafer1998,Schafer1999} as well as in sleeping humans \citep{Cysarz2004a,Bartsch2007}. A high degree of synchronization was reported for subjects during meditation with very little coordination seen during spontaneous breathing \citep{Cysarz2005}.  Raschke et al. \citep{Raschke1987} suggested that coordination between the cardiac and respiratory systems would be at its strongest during states of relaxation and reported strong coordination between the cardiorespiratory subsystems during sleep, \citep{Reed2007}, also stating that this coordination was easily disturbed under conditions of stress or disease.  Kabir et al. showed a reduction in phase coupling in patients with severe obstructive sleep apnoea (OSA) compared with mild OSA, synchronization levels also seemed to correlate with sleep stages  \citep{Kabir2010}.

Although neither the underlying mechanisms governing the coordination nor the physiological significance of such results is understood, its quantification could prove to have clinical merit, e.g. estimating the prognosis of cardiac diseases in patients having suffered myocardial infarctions \citep{Hoyer2002,Leder2000}.
%%%%%%%%%%%%

In  this study, we nvestigate  the  effects  of  extreme physical stress  on  cardiorespiratory synchronizations using the concept of phase locking with synchrogram and Empirical Mode Decomposition (EMD) analysis. 

We apply synchrogram analysis and EMD to respiration (RR)  and electrocardiogram (ECG) time series in order to find a mode that encapsulates the key features of the original signal.  The phase of this mode is calculated via the Hilbert transform and is compared with the phases from all modes of the corresponding ECG signal, after which the synchronization analysis is carried out. Specifically, we analyse the RR and ECG signals of Ironman athletes before and after the Ironmen race, when a Stroop test is administered. The ECG and  RR data are taken before and after  the athletes  perform  a Stroop test.  Our results consistently illustrated a rise in synchronizations after the  competition. Furthermore, we evaluate the control effect on synchronizations, we expect to see an increase in synchronizations between the cardiorespiratory systems after the race  due to an increased breathing rate, to which the heart adjusts its rhythm to beat at an equal rate \citep{Pokrovskii1985}. In both scenarios cardiorespiratory systems are trying to maintain homeostasis.
%%%%%%%%%%%%%%%%%%%%

The paper is organised as follows. Section ~\ref{Experiment} introduces the experimental settings  and data collection methods. Section ~\ref{Methods} considers the techniques applied for analysing the cardio- and respiratory time series data , followed by the results and discussion in Section ~\ref{Results}  and final conclusions in Section ~\ref{Conclusions}.

%%%%%%%%%%%%%%%%%%%%%%%
%%%%%%%%%%%%%%%%%%%%%%%%%

\section{Experimental Design and Data}
\label{Experiment}
%%%%%%%%%%%%%%%%%%%%%%%

\subsection{Experimental Design}
%%%%%%%%%%%%%%%%%%%%%%
The study investigated 14 Ironmen athletes before and after the race. The physical performance of the athletes was judged by the synchronization of the cardio- and respiratory systems. This was measured by taking the ECG and RR signals and studying their synchronization using time series analysis.  Stroop test was administered before and after the race to induce cognitive stress. Stroop mistakes were counted as a measure of cognitive performance. 

%%%%%%%%%%%%%
Simultaneous ECG and  RR signals were recorded from all participants for two settings: one during a Stroop test before and one after the Ironman race.

%%%%%%%%%%%%%%%%%%%%%
To remove the noise, we applied a zero-phase filter to the time series  signals using an IIR filter. The IIR filter has an 4th order and the cutoff frequency is 0.4. 
%%%%%%%%%%%%%%%%%

\subsection{Stroop Test}
%%%%%%%%%%%%%%%%%%%%%%%%%%%%%%
Cognitive stress is known to affect the physiological functioning of the cardiovascular system suppressing heart rate variability (HRV)  \citep{Hansen2003,Wood2002}.  In physiology, HRV is the variation in time interval between heartbeats, measured by the variation in the beat-to-beat interval \citep{Hon65}.   Raschke et al. suggested that coordination between the cardiac and respiratory systems would be at its strongest during states of relaxation and stated that this coordination was easily disturbed under conditions of stress or disease \citep{Raschke1987}.

\begin{figure}[!htp]
\centering
\includegraphics[width = 0.8 \linewidth]{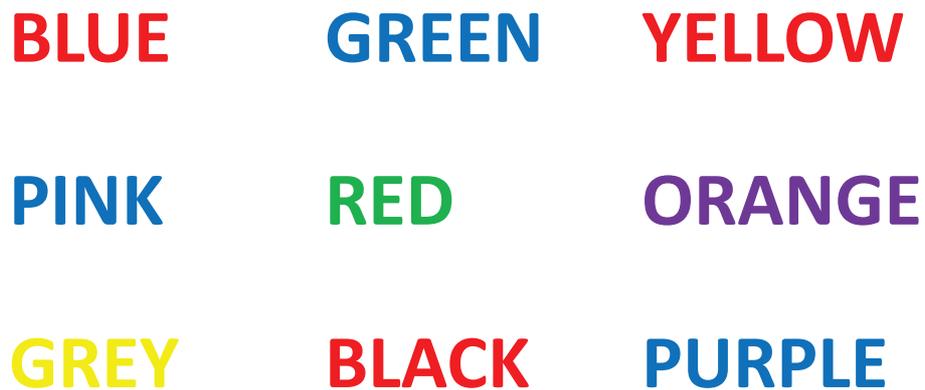}
\caption{Example of a Stroop test. Participants were required to state the colour of the word instead of reading the word.  For example the first line would be red, blue, red.}
\label{examplestroop}
\end{figure}

The Stroop effect is a demonstration of interference in the reaction time of a test \citep{Stroop1935}. Essentially, the name of a colour, e.g. “red”, is printed in a colour not denoted by the name. The example  in Fig.~\ref{examplestroop} shows  the word “blue”   printed in the colour red, and the word "red"  printed in colour "green". Naming the colour of the word takes longer and is more prone to errors than when the colour of the word and the name of the colour match. The Stroop test was applied in this study in order to turn the participants' attention away from consciously controlling their respiration depth and rate and instead to focus on completing the task at hand. In doing this, the unconscious, homeostatic mechanisms can be investigated and their influence on cardiorespiratory synchronizations found.

%%%%%%%%%%%%
%%%%%%%%%%%%

In this study,  the participants were asked in the first stage to complete a Stroop test - in order to impose stress and draw attention away from consciously controlling one’s breathing and instead focus on completing the task – pre Ironman event; and in the second stage they completed a Stroop test post Ironman event. 

%%%%%%%%%%%
We expect to see a decrease in the amount of synchronization during the Stroop test, due to the extreme physical stress.
%%%%%%%%%%%%%%%%%%%%%

\subsection{Data}
\label{Data}
%%%%%%%%%%%
%%%%%%%%%%%%%%
ECG was measured with three electrodes, positioned in Einthoven's triangle configuration,  and  recorded at 1000Hz. ECG and RR time series were recorded continuously using AcqKnowledge software (version 2.1). The signals were pre-processed using Matlab in order to extract the R-peaks. As the  time series were noisy and strongly non-stationary, EMD was implemented to decompose and reconstruct the respiration signal free of noise. The respiratory signal was recorded  via a force transducer fixed to a belt around the chest.  Subjects were asked to expel air from their lungs as the transducer was first fit, and then were instructed to breathe normally. ECG and RR signals were recorded simultaneously for 6 minutes - 1 minute prior to a Stroop test and five minutes during the test. The descriptive statistics of each individual in the study is   given in  Table~\ref{stats}. All individuals had to perform two Stroop tests, one before and one after the Ironman event.
%BRING THIS BACK _ WITH THE DATA IN THE TABLE 1. It also includes the average heart and respiration rates pre- and post-Stroop test.
%%%%%%%%%%%%%%%%%%
%%%%%%%%%%%%%%%%%%%%%%%%%%%%%%%%%%
%%%%%%%%%%%%%%%%%%%%%%%%%%%%%%%%%

\begin{table}[ht]
 %%%%%%%%%%%%%
\label{interventionsstats}
\caption{Descriptive statistics of 14 Ironman athletes. Means and standard deviations are shown in the bottom rows.}
\begin{center}
\small {
%%%%%%%%%%%%%%%%%%%
\begin{tabular}{|c|c|c|c|c|c|c|c|c|c|c|}
\hline
ID & Age & Gender & \makecell{Height\\(m)} & \makecell{Weight\\(kg)} & BMI & Fit (h/wk) & \makecell{Race Time\\(mins)} & \makecell{Recovery\\(mins)}\\
\hline
1 & 36  & M  & 1.85  & 74 &  21.6 &  14 & 621 & 72  \\
\hline
2 &  43 &  M &  1.79 & 71 & 22.2  & 17 &  638 & 75  \\
\hline
3 &  &  M &  &  &  &  & 675 & 97   \\
\hline
4 &  35 &  M &  1.72 & 66 & 22.3 & 20 & 695 & 59  \\
\hline
5 &  30 &  M &  1.83 & 69 & 20.6 & 15 & 712 & 106   \\
\hline
6 &  40 &  M &  1.68 & 72 & 25.5 & 19 & 738 & 93   \\
\hline
7 &  44 &  M &  1.92 & 92 & 25   & 20 & 759 & 70  \\
\hline
8 &  35 &  M &  1.74 & 68 & 22.5 &  8 & 769 & 98  \\
\hline
9 &  43 &  M &  1.7 &  72 & 24.9 &  13 & 775 & 112   \\
\hline
10 &  21 & M &  1.91 & 79 & 21.7 &  9 & 784 & 200  \\
\hline
11 &  31 & F &  1.65 & 58 & 21.3 &  15 & 803 & 109   \\
\hline
12 &  30 & M &  1.84 & 75 & 22.2 &  15 & 809 & 120  \\
\hline
13 &  39 & M &  1.77 & 73 & 23.3 &  15 & 870 & 132  \\
\hline
14 &  21 & M &  1.82 & 67 & 20.2 &  25 & 948 & 142   \\
\hline
\hline
Mean &  34.5 &  {} &  1.79 &  72 &  22.55 &  15.77 & 756.85 & 106.04 \\
\hline
STD & 7.67 & {} & 0.09 & 7.88 & 1.67 & 4.59 & 88.12 & 36.19  \\
\hline
Med & 35 & {} & 1.81 & 72 & 22.16 &15  & 764 & 102 \\
\hline
\end{tabular}
}
\end{center}
\label{stats}
\end{table}
% %%%%%%%%%%%%%%%%%%%%

\section{Methods}
\label{Methods}
%%%%%%%%%%%%%%%%%%%%%%
\subsection{Descriptive Statistics}
Statistical analysis was performed on the  data for 14 Ironmen using the  package R. The arithmetic mean (mean), standard deviation (STD)  and the median (Med) of the eight variables: Age in years, Gender, Height in meters, Weight in kilograms, Body Mass Index (BMI), Fitness(Fit) in hours per week, Race Time in minutes and Recovery time in minutes are given in Table~\ref{stats}. 

\begin{figure}[!htp]
\centering
\includegraphics[width=.4\textwidth]{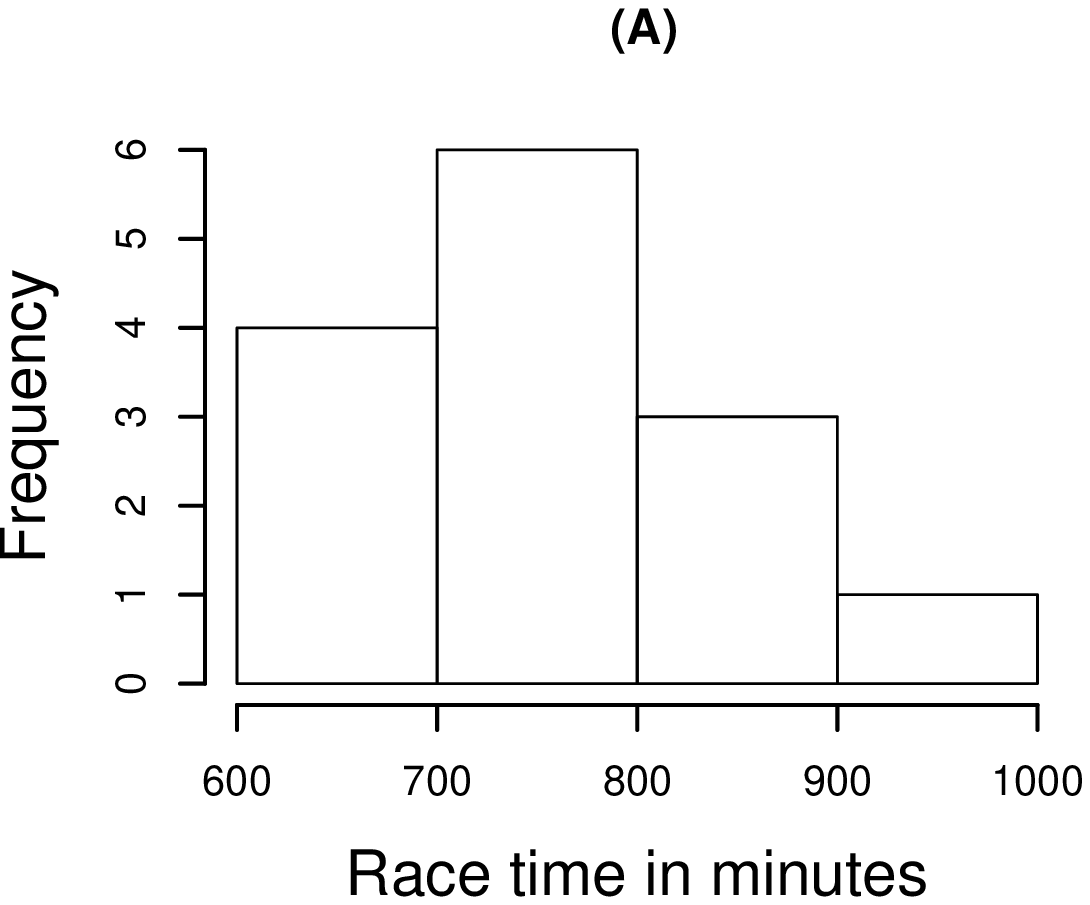}
\includegraphics[width=.4\textwidth]{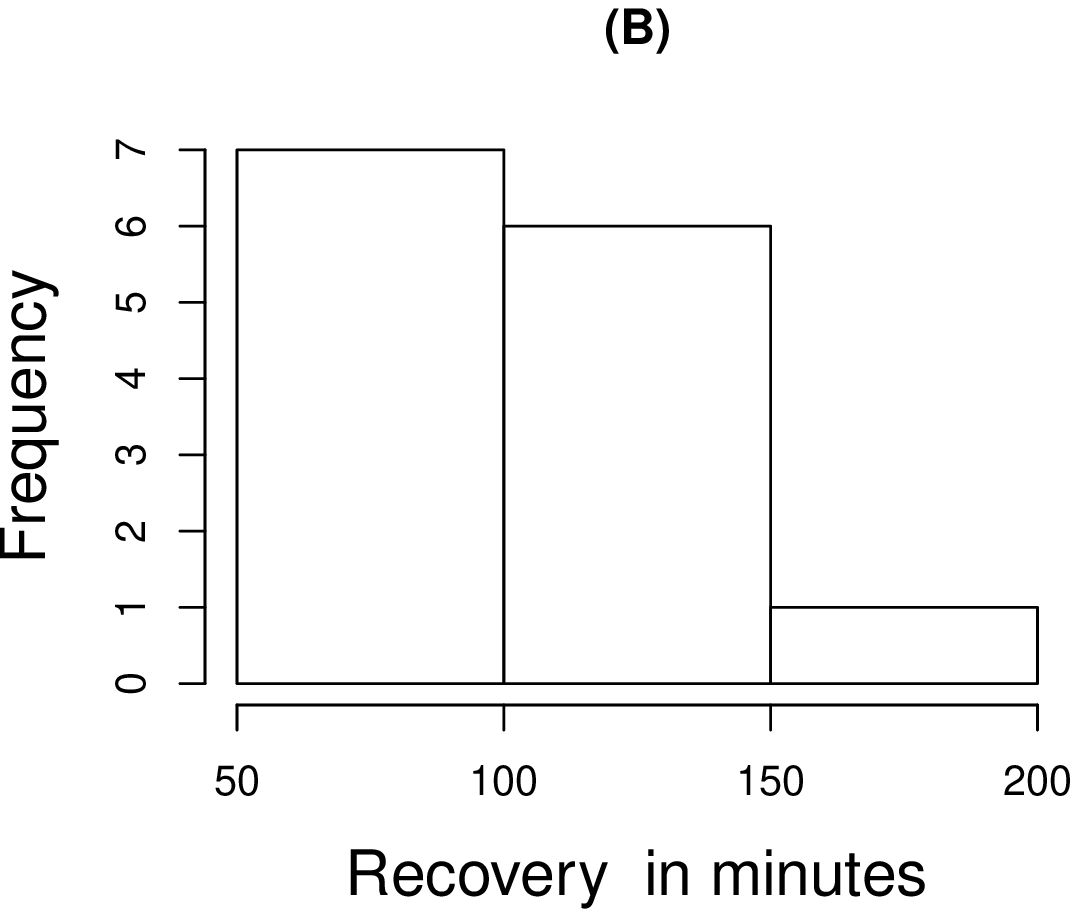}
\includegraphics[width=.4\textwidth]{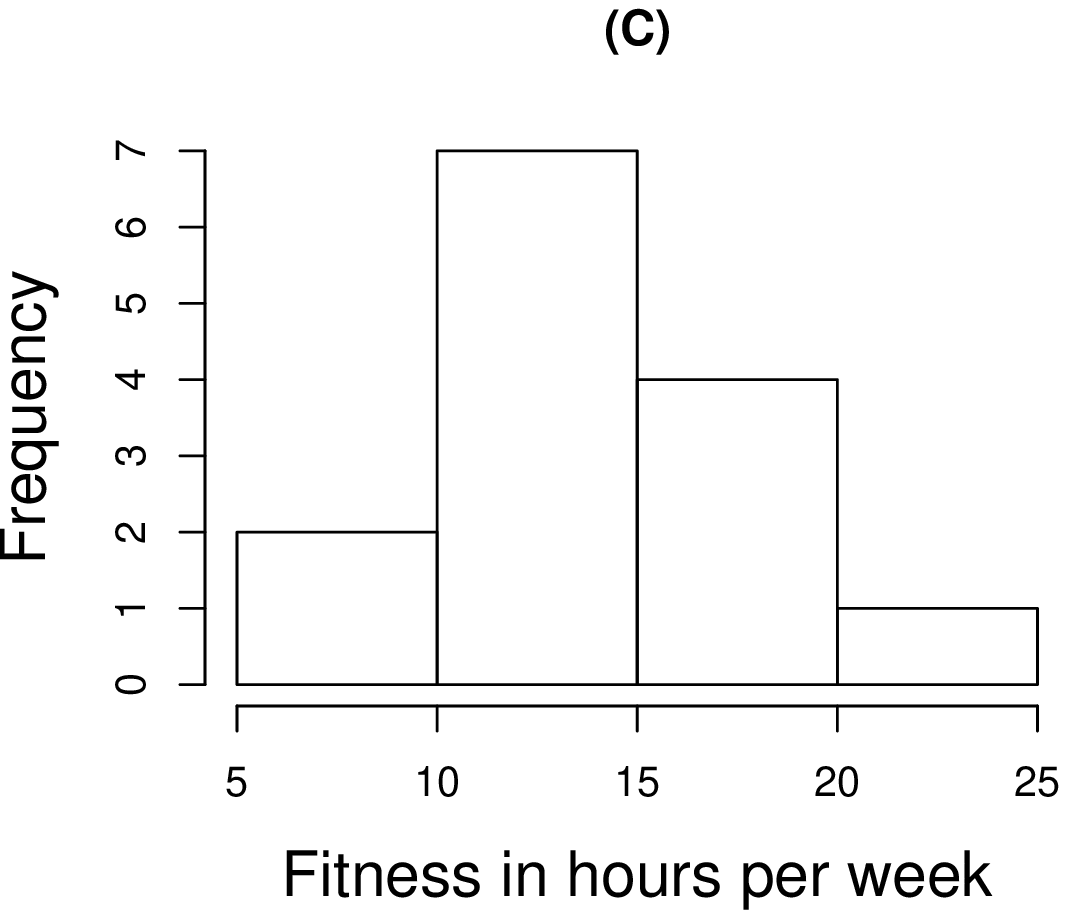}
\includegraphics[width=.4\textwidth]{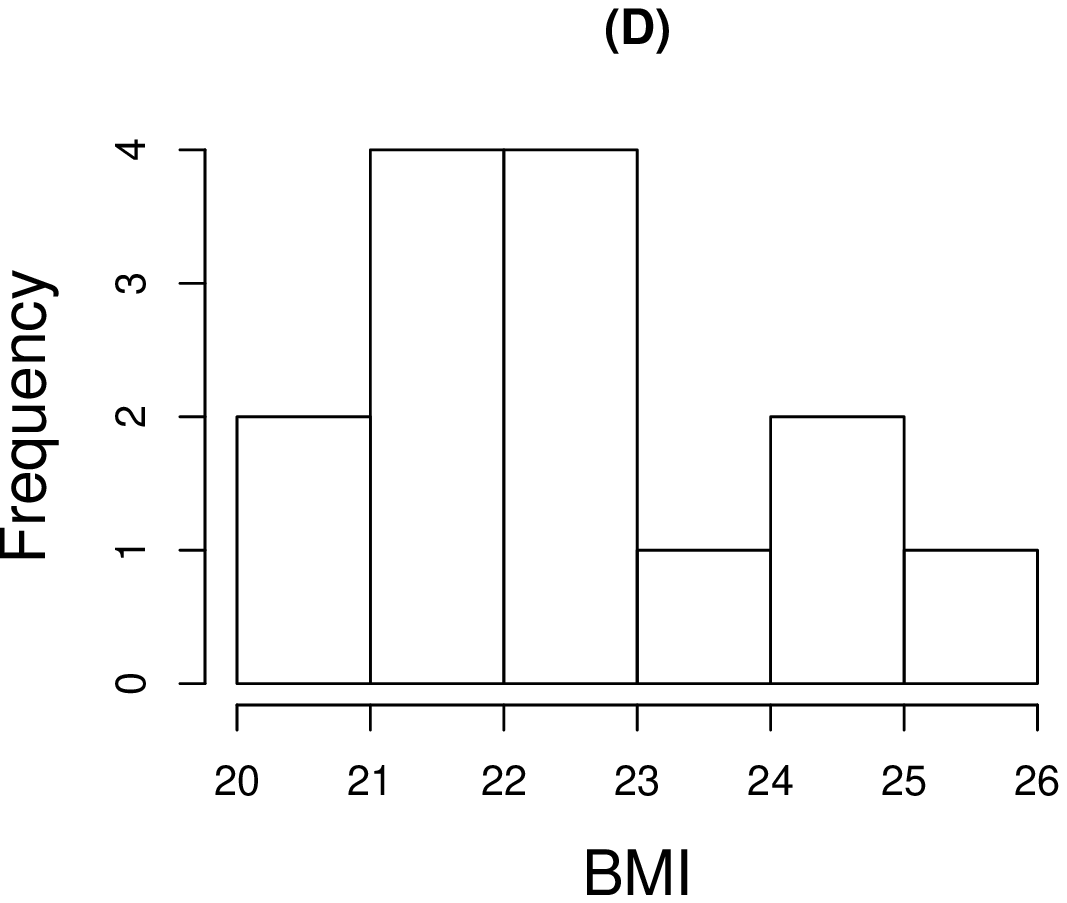}
\caption{Histograms of (A) Race and (B) Recovery time, (C) Fitness and (D) BMI for the athletes.}
\label{fig:Hist}
\end{figure}

Pearson correlation coefficients were computed for all pairs of variables. There was no linear correlation between variables, except some correlation between Race time and Recovery (0.6198), Age and Recovery (-0.6998), and Age and BMI (0.7949). The distribution of Recovery shows that only two athletes have a Recovery time significantly higher than the average - \#10 almost double the average and \#14 almost 50\% above the average. This indicates  that the majority of the  athletes are recovering in a similar way from the extreme race. While the race time for athlete \#10 is just above the average, the race time for  \#14  is way above the average. The Fit for \#14 is significantly above the average while the same variable for \#10 is below the average. The remaining variables for these two athletes  are in line with those of the remaining 12 athletes. The histograms of Race and Recovery time,  Fit and BMI are given on Figure~\ref{fig:Hist}.

As the sample size is small, we will focus the analysis on the signals measured. For each athlete we have 5 minutes of each ECG and RR signals measured twice, namely during the Stroop tasks completed before and after the competition.  This gives us a sufficiently large sample size to analyse the signals using advanced methods of signal processing (EMD) and synchrograms. Traditional statistical methods are not appropriate for the signal processing due to the amount of noise and complexity of the signals.

\subsection{Synchronizations}

The synchronization is a basic phenomenon in nature \citep{Rosenblum1996,Pikovsky1997,Cysarz2004b, Rosenblum2004}.  
Through the detection of synchronous states we may be able to achieve a better understanding of physiological functioning.
In the classical sense of periodic, self-sustained oscillators, synchronization is usually defined as the locking (entrainment) of the phases with a near constant phase difference that persists over time:
\begin{equation} \label{sync}
\phi_{n,m}=n\Phi_{1}-m\Phi_{2}=const
\end{equation}
where $n$ and $m$ are integers, $\Phi_{1},\Phi_{2}$ are phases of the two oscillators and $\phi_{n,m}$ is the generalized phase difference \citep{Tass1998}. In such cases, the $n:m$ phase locking demonstrates itself as a variation of $\phi_{n,m}$ around a horizontal plateau. We will use the length of this plateau as a measure of synchronization.

The phase $\phi(t)$ is  easily estimated from any scalar time series. A problem arises if the signal contains multiple component or time-varying spectra, thus making phase estimation difficult.  The EMD method overcomes this as it breaks a signal down into a finite set of components for which the instantaneous phase can be defined. 

As with most physiological time series, when investigating the coupling of the cardiorespiratory systems, noise can occur.  This noise originates not only from measurements and external disturbances, but also from the fact that there are other subsystems that take part in the cardiovascular control \citep{Stefanovska1999}.  These influences, when doing synchronization analysis, are also considered as noise.
%%%%%%%%%%%%%%%%%%%%%%%%
%%%%%%%%%%%

\subsection{Hilbert-Huang Transform}

%%%%%%%%%%%%
To study the phase synchronization  of the cardiorespiratory system we use Hilbert-Huang Transform (HHT) \citep{Huang1998,huang2005hilbert,Huang2008}. It is superior to the Fourier-based methods, which are the simplest and most popular methods of decomposing a signal into energy-frequency distributions. The  Fourier methods lose track of time-localised events and are proven ineffective when analysing physiological systems with non-stationary processes.  A popular alternative to Fourier methods is wavelet analysis.  It overcomes problems with non--stationarity, however, due to the use of single, basic wavelet is non-adaptive and therefore needs to be applied with care to non-linear data.
HHT is used in order to analyse non-linear and noisy signals as it describes them more locally in time. HHT is also capable of measuring instantaneous frequency and phase, which makes it particularly suitable for physiological time series. Hilbert-Huang transform applies Hilbert transform to intrinsic mode functions obtained from the EMD decomposed signals.

In the standard Hilbert Transform (HT),  $y_i$, can be written for any function $x_i$ as follows,
%%%%%%%%%%
\begin{equation} \label{Hilbert transform}
y_i=\frac{1}{\pi} P \int_{-\infty}^{\infty} \frac{x_{i}(t^{'})}{t-t^{'}}dt^{'},
\end{equation}
%%%%%%%%%%
where $P$ indicates the Cauchy principal value.
Gabor et al. determined that an analytical function can be formed with the HT pair \citep{Gabor1946},
%%%%%%%%%%%
\begin{equation} \label{HTanalytical}
z_{i}(t)=x_{i}(t)+{\it i}y_{i}(t) \equiv A_{i}(t)e^{{\it i}\phi_{i}(t)},
\end{equation}
%%%%%%%%
with amplitude $A_{i}(t)$ and  instantaneous phase $\phi_{i}(t)$,
\begin{equation}
A_{i}(t)=\sqrt{x_{i}^{2}(t)+y_{i}^{2}(t)},
\end{equation}
%%%%%%%%%%%%%
%%%%%%%%%%%
\begin{equation} \label{phase_0}
\phi_{i}(t)=\tan^{-1}\Big( \frac{y_{i}(t)}{x_{i}(t)} \Big).
\end{equation}
The instantaneous frequency can be presented as the time derivative of the phase,
\begin{equation}
\omega = \frac{d\phi_{i}(t)}{dt}.
\end{equation}
%%%%%%%%%%%%
When determining  the instantaneous phase, an assumption is made that the system studied can be modelled as weakly-coupled oscillators \citep{Stefanovska1999}. We also assume that their interactions can be investigated by analysing such phases \citep{Kuramoto1984}. We should note that the Hilbert transform is not the only method to estimate phase relationships, this can also be done by using wavelet transform or marked events methods \citep{Stefanovska1999,LeVanQuyen2001,Clemson2014}. Another main advantage of the HT is that it can find the phase of a single oscillation directly.
%%%%%%%%%%
%%%%%%%%%%%%%%%%%%%%%%%%%

\subsection{Synchrograms}

%%%%%%%%%%%%
In 1998, Schafer et al. developed the cardiorespiratory synchrogram in order to analyse $n:m$ synchronizations in the cardiorespiratory systems, where the heart beats $n$ times in $m$ respiratory cycles \citep{Schafer1998,Schafer1999}. The synchrogram analysis is very effective to study phase synchronization between a point process (heartbeat) and a continuous signal (respiration).  This technique has been used to look at synchronizations in infants \citep{Mrowka2000}, in adults during poetry recitation \citep{Cysarz2004b} and desynchronizations following myocardial infarction \citep{Leder2000}.  A high degree of synchronization was reported for subjects during meditation with very little coordination seen during spontaneous breathing \citep{Cysarz2005}. Bartsch et al \citep{Bartsch2012} used the synchrogram method to investigate the response of cardiorespiratory synchronization to changes in physiological states through sleep.

\begin{figure}[!htp]
\includegraphics[width=\linewidth]{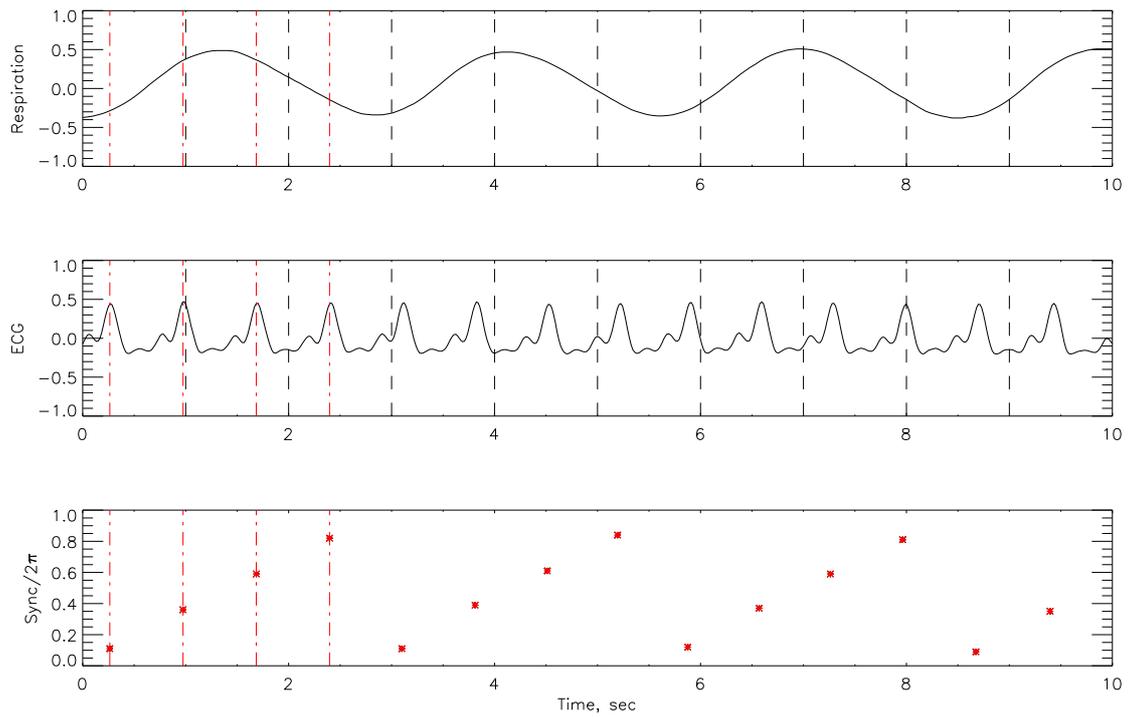}
\caption{An example of how the cardiorespiratory synchrogram works.  On the top is the respiration, in the middle  the corresponding ECG signal and at the bottom, the formation of the synchrogram. The position of each heartbeat in relation to its appearance in the phase of the respiratory cycles can be clearly seen. Red broken vertical lines indicate picks in the heart beats and black vertical broken lines indicate one second interval. This  example illustrates $n:1$ synchronization.}
\label{synceg}
\end{figure}

%%%%%%%%%%%%%%%%%%%%%%%
After cleaning the signal with a low pass  filter,   
Matlab code was employed to  the respiratory signal, to detect R-peaks from the ECG time-series. The Hilbert transform was used to calculate the instantaneous phase of the respiration signal $\Phi_{nr}$ from (\ref{phase}).   We then considered the respiratory phase at times $t_k$ - the $r$-peak of the $k^{th}$ heartbeat. The cardiorespiratory synchrogram can be constructed by observing the phase of the respiration at each $t_k$, and wrapping the phase into a $[0, 2{\pi}m]$ interval. In the simplest case of $n:1$ synchronization, there are $n$ heartbeats in each respiratory cycle.  Plotting these relative phases $\Psi_{n,1}$ as a function of time against $t_k$, we  observe $n$ horizontal lines (representing the number of heartbeats) in one  respiratory cycle. This is illustrated in Fig.~\ref{synceg}. The relative phase is given by,
%%%%%%%%%%%
\begin{equation} \label{psi}
\Psi_{n,m}(t_k)=\frac{1}{2\pi}[\Phi_{nr}(t_k)\mod 2\pi{m}].
\end{equation}
%%%%%%%%%%%
%%%%%%%%%%

%%%%%%%%%%%%%%%%%%%%%%%
\subsection{Empirical Mode Decomposition}

As it has been noted above, HHT consists of two stages in order yo analyse a time series.  The first stage, EMD, decomposes a time series into a set of simple oscillatory functions, defined as intrinsic mode functions (IMFs).   Typically, an IMF is a function that fulfills the following:
\begin{itemize}
\item In the entire dataset, the number of extrema and the number of zero-crossings must be either equal or differ by at most one
\item At any point, have a mean value of zero between its local maxima and minima envelopes.
\end{itemize}
The IMF components are obtained by applying an iterative technique known as 'sifting', this process is as follows:

\begin{enumerate}
\item Localize all the local maxima in the time series ($x(t)$) and connect them with a cubic spline, this is the upper envelope. Repeat the procedure with the local minima defining the lower envelope.

\item Calculate the mean of the upper and lower envelopes $m_{1}(t)$ and determine the first component by subtracting the mean from the original time series $x(t)$.

\begin{equation}
c_{1}(t)=x(t)-m_{1}(t)
\end{equation}

\begin{itemize}
\item if the condition for an IMF are met then the component $c_{1}(t)$ is an IMF.
\item if the conditions are not met, repeat the process from step 1 until an IMF is found.
\end{itemize}

\item Finally, subtract the IMF component from the original time series to find the residue, $r_{1}(t)$.

\begin{equation}
r_{1}(t)=x(t)-c_{1}(t)
\end{equation}

\item Repeat the sifting process using $r_{1}(t)$ as the new time series.

\item Continue this process until all of the intrinsic modes ($c_{i}$) are found.
This process can be terminated when the $n$th residue is a monotonic function that doesn't present any extrema and no more IMFs can be extracted. 
This last residue is called the trend of the data. It is important to note that any residue constitutes a trend for  the previously extracted oscillation. i.e. $r_{i}$ is the trend followed by the $c_{i}$ oscillation. 
\end{enumerate}

After this procedure it is possible to express the original data in terms of the obtained IMFs,

\begin{equation}
x(t)=\displaystyle \sum_{i=1}^{N}c_{i}(t)+r_{n}(t)
\label{eqc}
\end{equation}

Orthogonality of the EMD is not guaranteed theoretically, but is satisfied in a practical sense as the IMFs are orthogonal within a certain period of time. In this sense the process only ensures time localized orthogonality.

The instantaneous phase can be calculated by applying the Hilbert transform to each IMF, $c_{i}(t)$. The procedures of the Hilbert transform consist of calculation of the conjugate pair of $c_{i}(t)$, i.e., 
\begin{equation}
y_i=\frac{1}{\pi} P \int_{-\infty}^{\infty} \frac{c_{i}(t^{'})}{t-t^{'}}dt^{'},
\end{equation}
 where $P$, as in equation (\ref{Hilbert transform}), indicates the Cauchy principal value. With this definition, two functions $c_{i}(t)$ and $y_{i}(t)$ forming a complex conjugate pair, define an analytic signal $z_{i}(t)$:
\begin{equation}
z_{i}(t)=c_{i}(t)+{\it i}y_{i}(t) \equiv A_{i}(t)e^{{\it i}\phi_{i}(t)},
\end{equation}
with amplitude $A_{i}(t)$ and the instantaneous phase $\phi_{i}(t)$:
\begin{equation}
A_{i}(t)=\sqrt{c_{i}^{2}(t)+y_{i}^{2}(t)},
\end{equation}
\begin{equation} \label{phase}
\phi_{i}(t)=\tan^{-1}\Big( \frac{y_{i}(t)}{c_{i}(t)} \Big).
\end{equation}

\begin{figure}[!htp]
\includegraphics[width= 0.99\textwidth, height=0.80\textwidth]{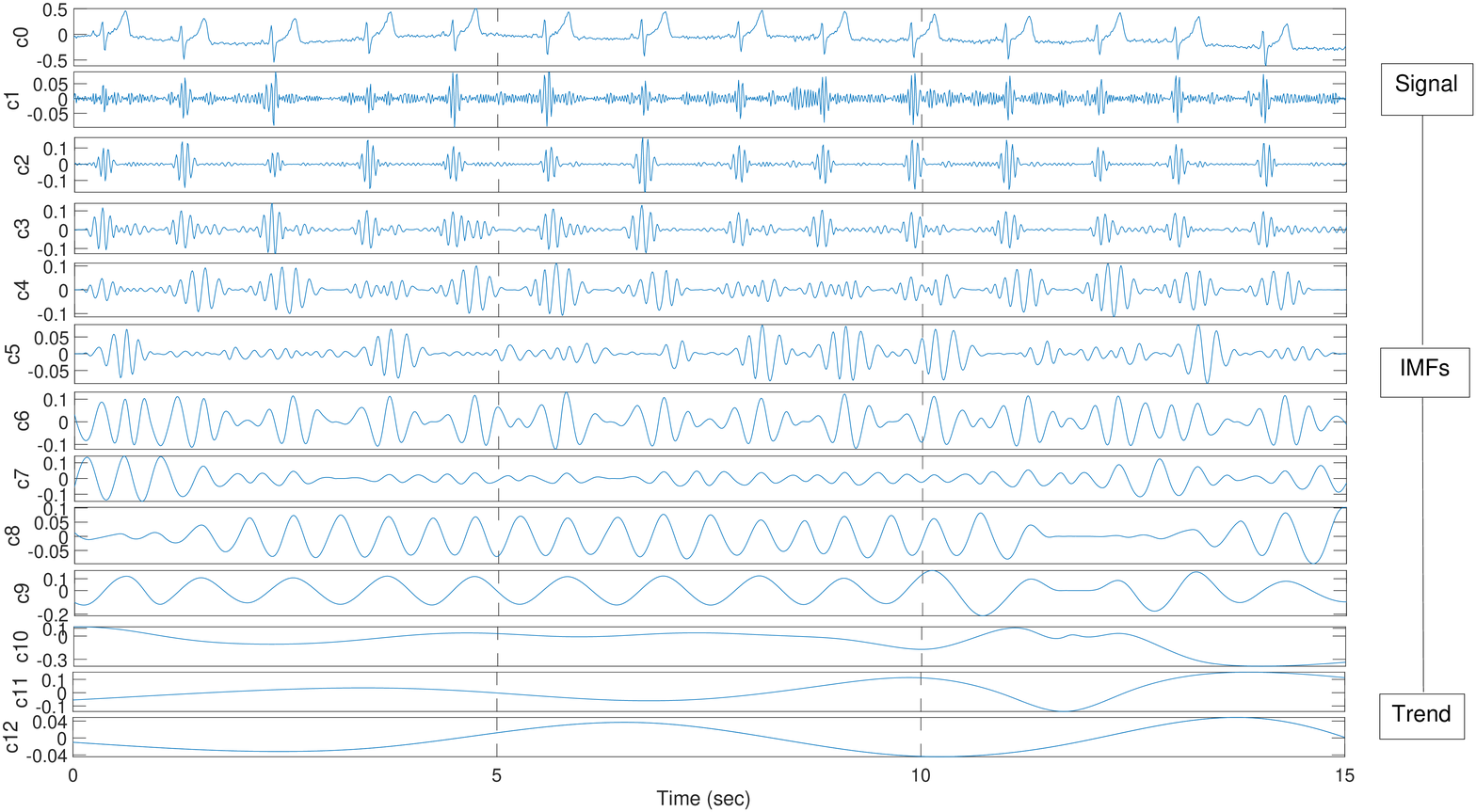}
\caption{Illustration of EMD decomposition into IMFs for athlete \#14.}
\label{IMF}
\end{figure}

\begin{figure}[!htp]
\includegraphics[width=0.99\linewidth]{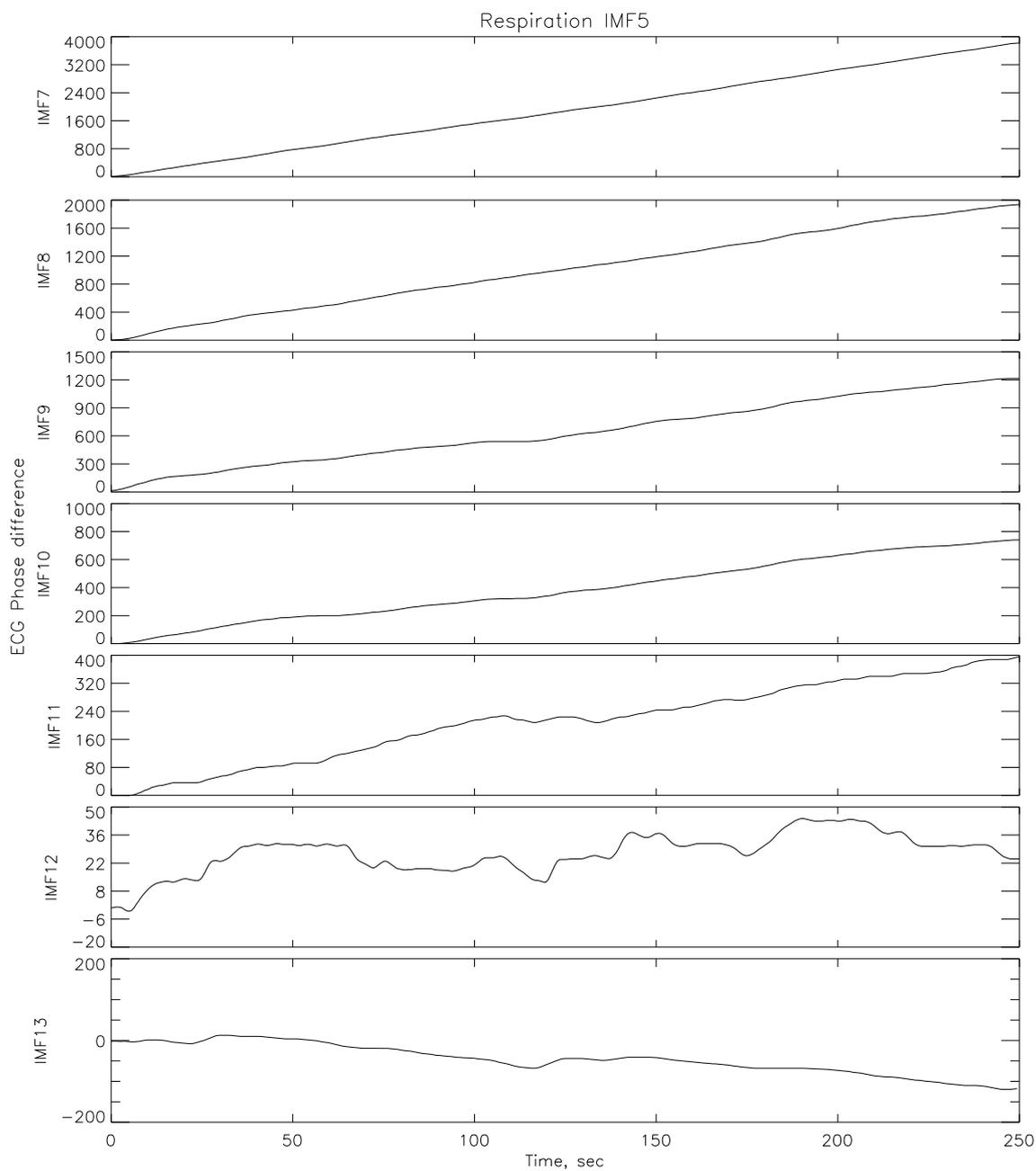}
\caption{Phase differences between one IMF (IMF5) from respiration decomposition and several IMFs from the ECG decomposition (athlete \# 14 post-race).}
\label{phasediffs}
\end{figure}

An illustration of EMD decomposition into EMFs is given on Fig.~\ref{IMF}. 
%%%%%%%%%%%%%%%
%%%%%%%%%%%%%%%%%
EMD was applied to corresponding ECG and RR time series in order to find the  synchronized modes.  The underlying theory is that in the decomposition of the ECG, lies an IMF (or set of IMFs) that describe the influence that respiration has on the dynamics of the heart.  Once the IMFs for both time series have been found, a particular IMF is found from the decomposition of the respiration signal which contains the key features of the original signal, while neglecting the faster oscillations as noise.
The decomposition in IMFs is illustrated on Fig. \ref{IMF}. The Hilbert transform is applied to this mode as well as all the other modes from the ECG signal and the instantaneous phases, $\phi_{i}(t)$ , are calculated with  (\ref{phase}).  A vector matrix is constructed showing the phase differences between the respiration IMF and all of the IMFs from the ECG decomposition (\ref{sync}) where areas of plateaus show synchronous periods between the cardiorespiratory systems, see Fig.~\ref{phasediffs}.  Here, the phase differences between IMF5 from the respiration signal and IMFs from the ECG signal were computed. For the practical purpose of this paper,
IMF5 was chosen as the base RR IMF to compute the differences with the ECG IMFs, as it just starts showing significant difference with the previous IMFs from the RR spectrum  \citep{huang2005hilbert}.
%12 from the ECG are clearly the most synchronised with long periods of plateauing compared to the other IMFs.
%\clearpage

%%%%%%%%%%%%%%%%%%%
%%%%%%%%%%%%%%%%%%%%%%%

\vspace{1cm}
%%%%%%%%%%%%%%%%%%%%%%%%%%%%%%%%%
\subsection{Data Analysis} 
The steps of data processing were done with as follows: 
\begin{enumerate}
    \item 
EMD was applied to corresponding ECG and RR signals taken from athletes performing a Stroop test both before and after the  Ironman competition. The number of sifting times depends on the data quality, and it varies case by case. Since the ECG and RR signals were pre-processed to a sampling rate of 100Hz, we set the sifting time as 100.
\item Visually the resulting IMFs decomposed by the EMD were inspected. If the amplitude of a certain model is dominant and the wave form is well distributed, the data are said to be well decomposed and the decomposition is successfully completed. Otherwise, the decomposition may be inappropriate, and we have to repeat step (1) with different parameters.
\end{enumerate}

%%%%%%%%%%%%%%%%%%%%%%%%

\section{Results}
\label{Results}
%%%%%%%%%%%%%%%%%%%%%%%%%%%%
%\subsection{Statistical Analysis}
%%%%%%%%%%%%%%%%%%%%%%
\subsection{Synchrograms}

%%%%%%%%%%%%%%%%%%%%%%%%%%%%%

\begin{figure}[h]
    \centering
    \includegraphics[width=\linewidth]{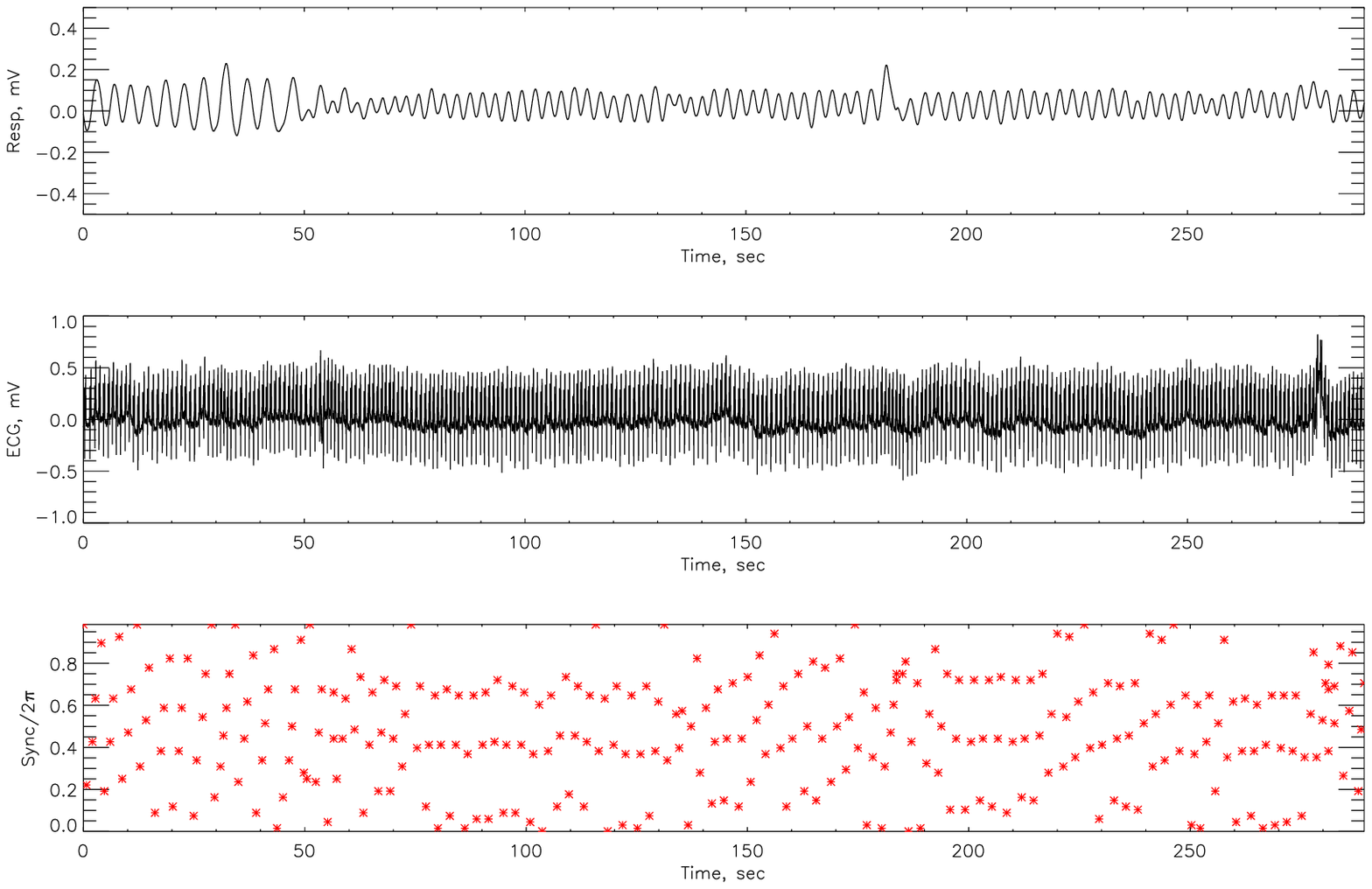}
    \caption{The cardiorespiratory synchrogram for an athlete \#14 completing a Stroop test before Ironman Competition, 3:1 phase locking.}
    \label{fig:falgun_pre}
\end{figure}

\begin{figure}[h]
    \centering
    \includegraphics[width=\linewidth]{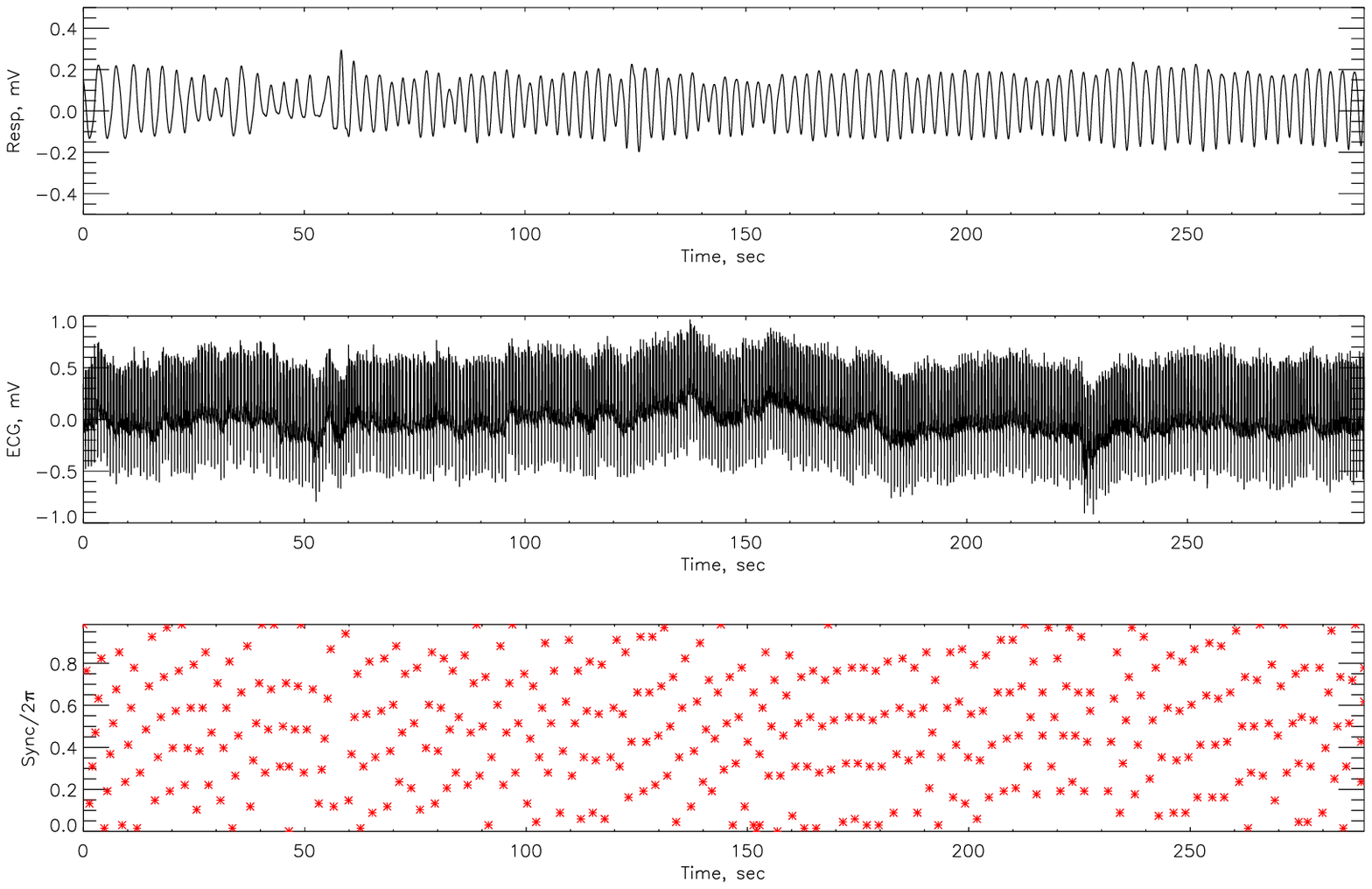}
    \caption{The cardiorespiratory synchrogram for an athlete \#14 completing a Stroop test after Ironman Competition, with 4:1 phase locking.}
    \label{fig:falgun_post}
\end{figure}

The cardiorespiratory synchrograms were calculated for each athlete  pre- and post-Ironman race for one and two respiration cycles, $ m=1,2$. Exemplary synchrograms for one athlete \#14 completing a Stroop test before and after competition are shown in Figures~\ref{fig:falgun_pre} and~\ref{fig:falgun_post}, respectively. The synchronization level for this athlete changes from 3:1 to 4:1. Figure~\ref{fig:falgun_pre} shows numerous regions of 3:1 phase locking where the heart beats 3 times for every respiratory cycle. In contrast, the synchrogram results after the competition for the same athlete on Figure~\ref{fig:falgun_post}, illustrate that not only have the regions of synchronization become longer and increased in stability, but the $n:m$ ratio has increased to 4:1. 

\begin{figure}[h]
    \centering
    \includegraphics[width=\linewidth]{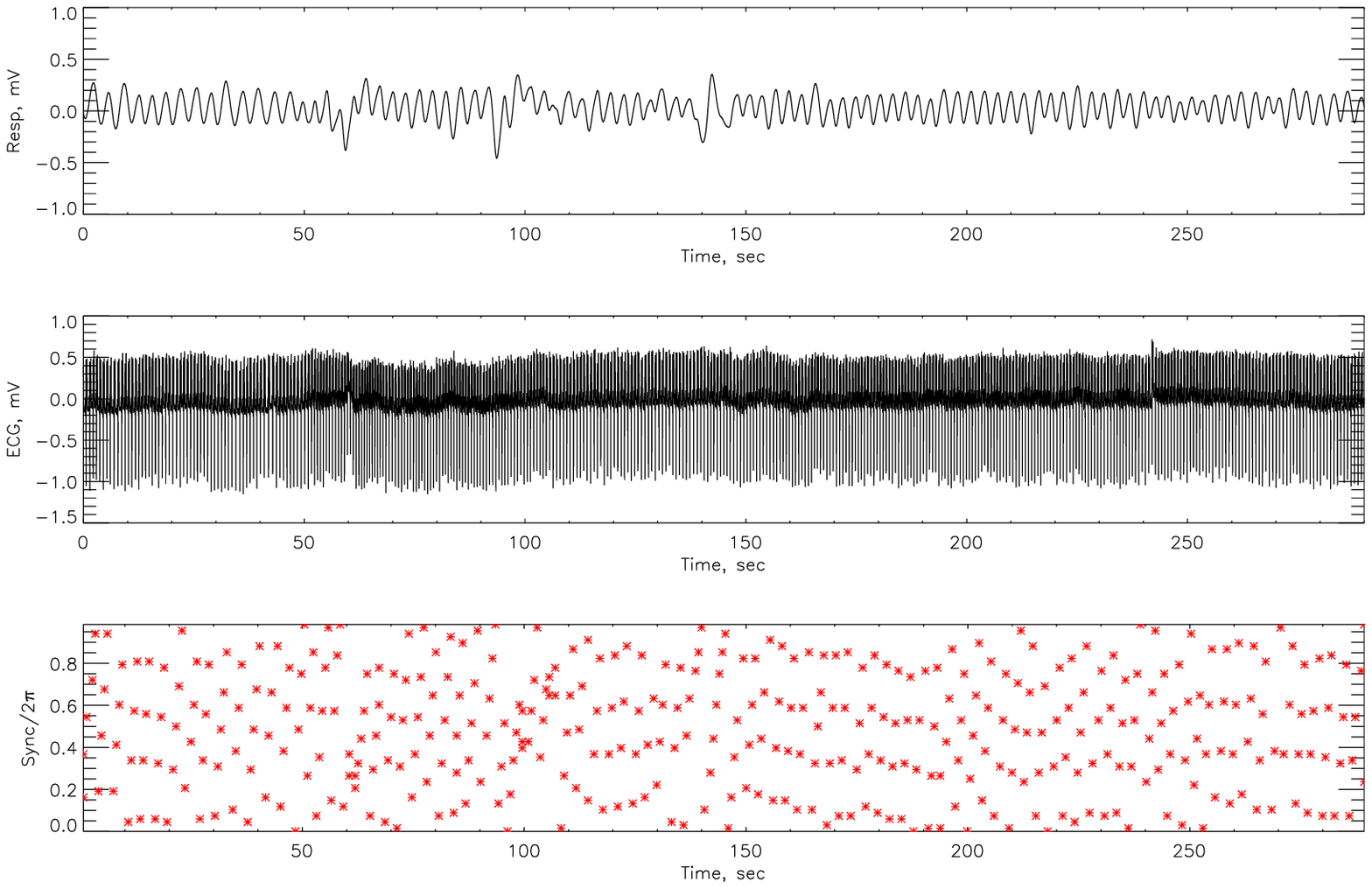}
    \caption{The cardiorespiratory synchrogram for an athlete \#3 completing a Stroop test before Ironman Competition, with 4:1 phase locking.}
    \label{fig:micheal_pre}
\end{figure}

\begin{figure}[h]
    \includegraphics[width=\linewidth]{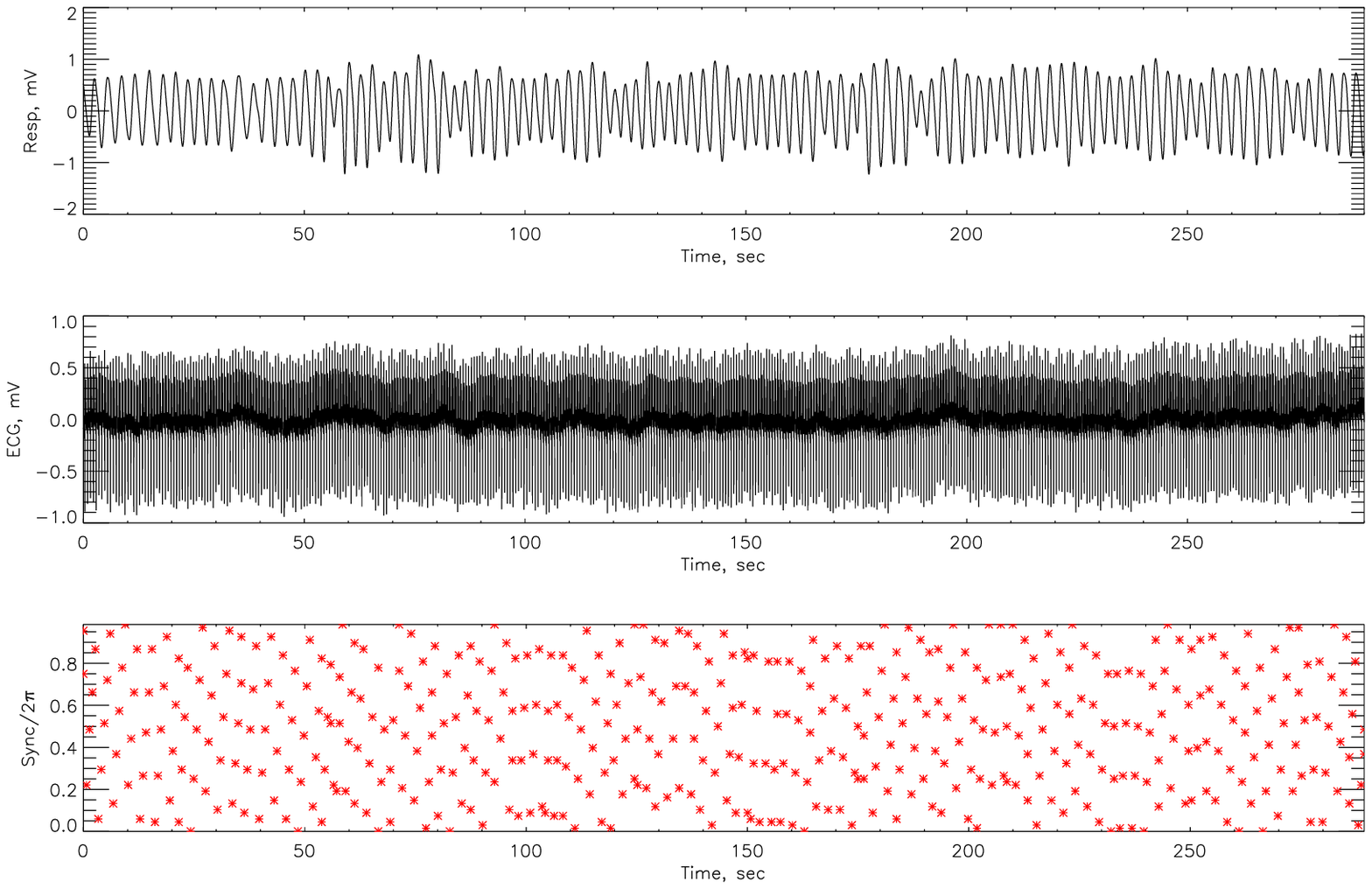}
    \caption{The cardiorespiratory synchrogram for an athlete \#3 completing a Stroop test after Ironman Competition, with 4:1 phase locking but longer time.}
    \label{fig:micheal_post}
\end{figure}

The synchrograms for athlete \#3, shown in Figures~\ref{fig:micheal_pre} and~\ref{fig:micheal_post}, support these findings with synchronization of 4:1 locking observed both pre- and  post-race. However, in  Figure~\ref{fig:micheal_pre} we see far more regions containing no synchronization at all. The regions of synchronization seen post-competition are with significantly increased length and of 4:1 locking.

%%%%%%%%%%%%%%%
%%%%%%%%%%%%%%%
%%%%%%%%%%%%%%%%%%%

%%%%%%%%%%%%%%%%%%%%%%%%%%
%%%%%%%%%%%%%%%%%
%%%%%%%%%%%%%%%%%%%%%%

%%%%%%%%%%%%%%%%%%%%%%%%%
The performance of the Ironman athletes in the Stroop test before and after the competition was not found to be different. This indicated that they were focused on the cognitive task and did not control their breathing. As the purpose of the test was to turn the participants' attention away from consciously controlling their respiration depth and rate and instead to focus on completing the task, we have not included in the analysis   the number of errors for each participant.
%%%%%%%%%%%%%%%%%%%%%%%

%%%%%%%%%%%%%%%%%%%%%%
\subsection{Phase Difference}
%%%%%%%%%%%%%%%%%%%%%%%%%%%%%%%%

%%%%HERE TAble 2
\begin{table}[h]
\caption{Summary of synchronization results for the Ironman athletes. The synchronization duration is shown for each of the athletes (rows) for each of the detected synchronization levels for pre- and post-Ironman Stroop tests. Also shown are the pre- and post-Ironman total duration of synchronization periods for each of the athletes (totals pre- and post- columns), and total duration for each of the synchronization levels pre- and post-Ironman.}
\begin{center}
\small{
%%%%%%%%%%%%%%%%%%%
\begin{tabular}{|c|c|c|c|c|c|c|c||c|c|c|c|c|c|c|c|}
\hline
ID & \multicolumn{7}{c||}{pre-Ironman} & \multicolumn{6}{c|}{post-Ironman} & \multicolumn{2}{c|}{Totals} \\
\hline
{} & 2:1 & 3:1 & 4:1 & 5:1 & 5:2 & 7:2 & 9:2 & 3:1 & 4:1 & 5:1 & 5:2 & 7:2 & 9:2 & pre & post \\
\hline
1 & -  & -  & -  & - &  - &  - & 50 & - & 60 & - & - & - & 250 & 50 & 310 \\
\hline
2 &  45 &  60 &  - & - & -  & - &  - & - & 60 & - & - & 120 &  -  & 105 & 180 \\
\hline
3 & - &  160 & - & - & - & - & - & - & 250 & - & - &  30 & -  &160 & 280 \\
\hline
4 & - & 90 & 20 & - & - & - & - & 30 & 150 & - & - & - & - & 110 & 180 \\
\hline
5 & - & - & 45 & - & - & 50 & - & 45 & 90 & - & - & 25 & - & 95 & 160 \\
\hline
6 & - & 30 & 40 & - & - & -  & 70 & 30 & 120 & - & - & - & - & 140 & 150 \\
\hline
7 & - & - & 180 & - & - & - & - & - & 220 & - & - & - & - & 180 & 220 \\
\hline
8 & - & 50 & 60  & - & - &  40 & - & 75 & - & - & - & 30 & - & 150 & 105 \\
\hline
9 & - & - & - & 60 & - & - & 20 & - & - & 90 & - & 130 & - & 80 & 220 \\
\hline
10 & - & 60 & - & - & - & - & - & 250 & - & - & 80 & - & - & 60 & 330 \\
\hline
11 & - & 90 & - & - & 60 & - & - & - & 80 & - & - & 90 & - & 150 & 170 \\
\hline
12 & - & - & 90 & - & - & - & - & - & 80 & - & 60 & - & - & 90 & 140 \\
\hline
13 & - & 120 & 40 & - & - & - &- & 200 & - & - & 30 & - & - & 160 & 230 \\
\hline
14 & - & 40 & - & - & - & - & 70 & 45 & 90 & - & - & - & - & 110 & 135 \\
\hline
Tot & 45 & 700 & 475 & 60 & 60 & 90 & 210 & 675 & 1200 & 90 & 170 & 425 & 250 & 1640 & 2810 \\
\hline
\end{tabular}
}
\end{center}
%\label{sync}
\label{synchronisationstats} %WAS interventionsstats SHELYAG
\end{table}

%%%%%%%HERE Table 3

\begin{table}
    \centering
    \small{
    \begin{tabular}{|c|c|c|c|c|c|c|}
    \hline
     Parameter & \multicolumn{2}{c|}{Pearson $r$} & \multicolumn{2}{c|}{Spearman $\rho$} & \multicolumn{2}{c|}{$p$-value} \\ \hline
               & Pre   &  Post &  Pre  &  Post & Pre  & Post \\ \hline
     Age       &  0.38 & -0.03 &  0.31 &  0.29 & 0.30 & 0.33 \\ \hline
     Height    & -0.27 &  0.46 & -0.30 &  0.34 & 0.32 & 0.26 \\ \hline
     Weight    &  0.07 &  0.44 & -0.23 &  0.53 & 0.46 & 0.06 \\ \hline
     BMI       &  0.39 &  0.05 &  0.39 &  0.13 & 0.18 & 0.67 \\ \hline
     Fit       &  0.26 & -0.30 &  0.41 & -0.24 & 0.17 & 0.44 \\ \hline
     Race time &  0.33 & -0.27 &  0.26 & -0.22 & 0.39 & 0.48 \\ \hline
     Recovery  & -0.27 &  0.28 & -0.19 & -0.03 & 0.53 & 0.91 \\ \hline
    \end{tabular}
    }
    \caption{Pearson and Spearman rank correlation coefficients between the duration of synchronization periods pre- and post-competition (Table~\ref{synchronisationstats}) and the individual parameters of the athletes (Table~\ref{stats}). $p$-values for Spearman correlations are also shown. Small $p$-values indicate significant correlations.}
    \label{correlations}
\end{table}

%%%%%%%%%%%%%%%

\begin{figure}[!htp]
\includegraphics[width=\linewidth]{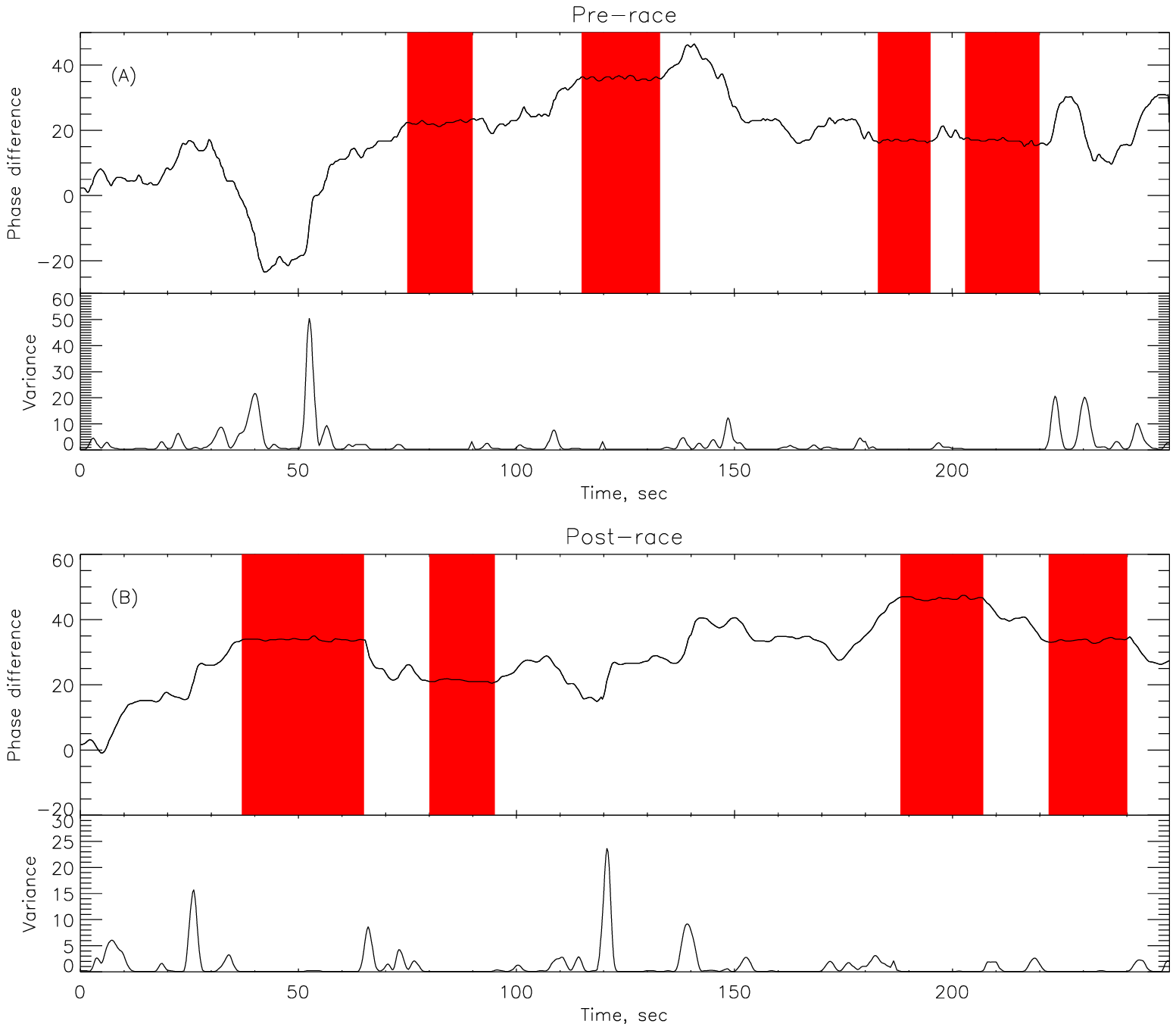}
\caption{EMD decomposition with data for athlete \#14 showing the plateau in pink for (A) Pre-race upper panel  and (B) Post-race upper panel. The length of synchronizations, given by the length of the plateaus (in pink) is larger in the post-race indicating that the athlete is more relaxed after the competition and shows a better synchronization between cardiac and respiratory systems. The lower  panel in (A) and (B) represents the variance in phase difference.  }
\label{test}
\end{figure}

Here we  compute  the phase difference  with $n:m$ as $n:1$ and $n:2$ for all athletes and using HHT to determine the instantaneous phase difference (\ref{phase_0}). The lengths of the plateaus, determined by the change of phase being a constant,  is a measure for each of the synchronization phase locking for both pre- and post-Ironman exercise are provided.
We computed the phase difference of ECG and respiration signals directly for both pre- and post-Ironmen test. Figure ~\ref{fig:falgun_pre} and ~\ref{fig:falgun_post} give the phase difference of participant \#14 before and after the competition. They show the phase of ECG is near constant in both pre- and post-competition, but the phase of respiration was improved. Meanwhile, the post-competition phase difference is smaller than that of the pre-competition one. For comparison, the results of athlete \#3 are given on Fig~\ref{fig:micheal_pre} and Fig~\ref{fig:micheal_post} for pre-and post-competition respectively. The synchrograms show that for Athlete \#3 the synchronization is stronger for post-competition. We analyzed this difference of all participants and found they all show the above phenomenon. In other words, the synchronization  between ECG and respiration signals appears to be stronger after the Ironmen competition. However, as the signals are  complex, the synchrograms are difficult  to read and interpret. This is rectified by using EMD with HHT phase locking as shown in Fig.~\ref{test} which represents athlete \#3 in the upper panels (a) pre- and (b) post-competition. The lower panels of (a) and (b) show the variance of the phase difference.
%%%%%%%%%%%%%%%%%
%%%%%%%%%%%%%%%%%%%%%

Table~\ref{synchronisationstats} illustrates synchronization results pre- and post-competition for all athletes, stating the duration of synchronization (in seconds) along with the level of synchronization. Shown are those levels $n:m$ where plateaus were observed for $n=2,3,...,9$ and $m=1,2$. Out of the 14 participants, only one (\#8) displayed longer regions of synchronization prior to the Ironman competition. Participant 10 exhibited one minute of 3:1 synchronization pre-competition but following the Ironman race had over 4 minutes of synchronization at the same level. 
The table clearly demonstrates that the synchronization is stronger (the synchronization periods are longer) for all except one (\#8) athletes. The difference is significant with the ratio of post- to pre-competition total synchronization periods of 1.7. Furthermore, an increase in synchronization for smaller synchronization levels 2:1 to 5:1 is 1.54, which is more than twice as small compared to the increase of 3.4 for synchronization levels 5:2, 7:2 and 9:2.

Further on, Pearson and Spearman rank correlation coefficients were computed between the duration of synchronization periods and individual parameters from Table~\ref{stats}. The results, together with the $p$-value for Spearman rank correlation are presented in Table~\ref{correlations}.

As is evident from the data presented in the table, there are no strong linear correlations between the parameters of the athletes and their synchronization times. The strongest Spearman correlation of 0.53 with the $p$-value of 0.06 is observed in weight - post-competition synchronization time parameter pair. Therefore, it can be concluded that physiological properties of each tested individual do not play a significant role in synchronization levels. It should be, however, noted that participating in Ironman competition already includes some implicit pre-selection.

\begin{figure}[!htp]
\includegraphics[width=\linewidth]{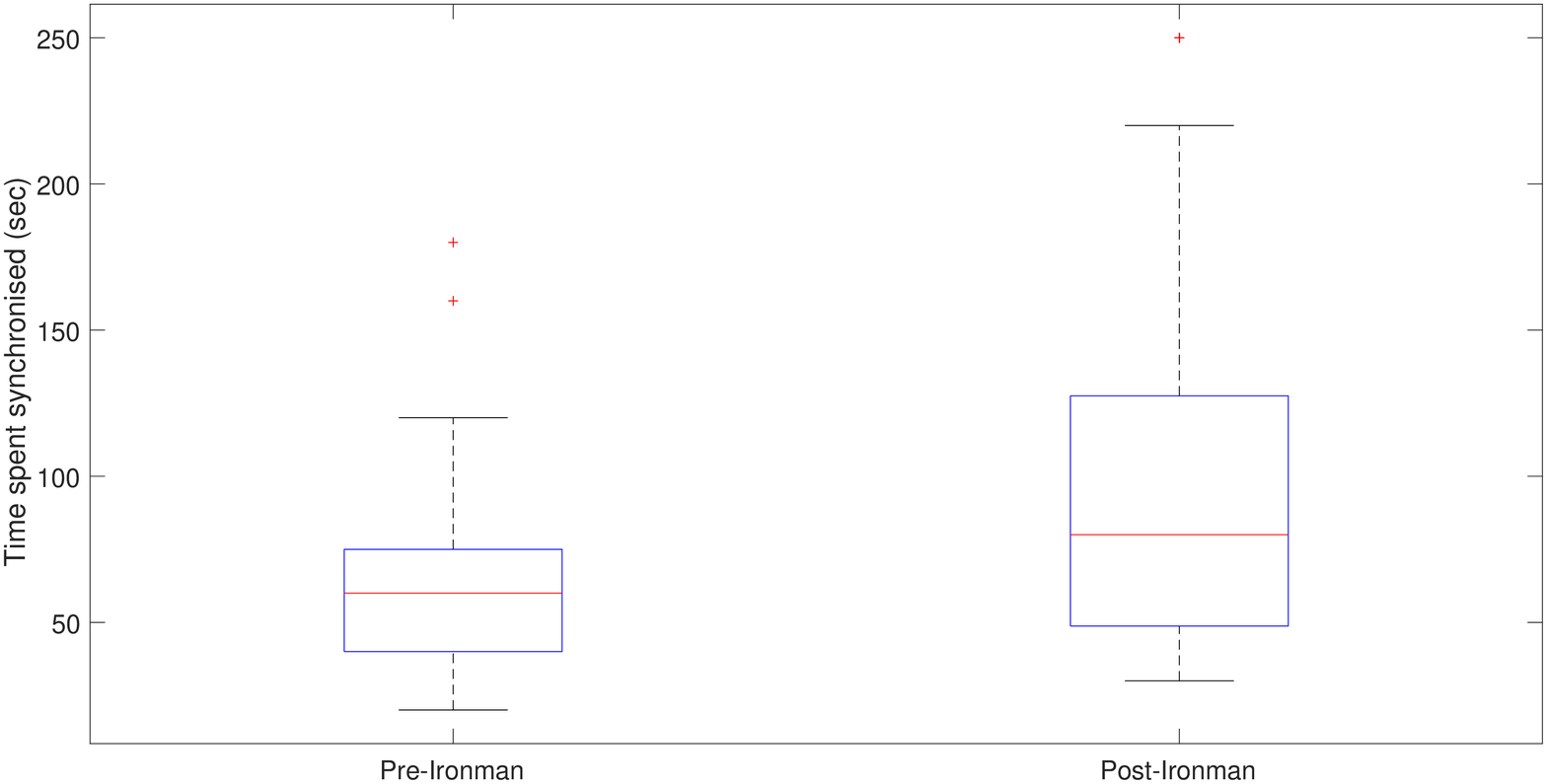}
\caption{Box plots displaying the times the cardiorespiratory systems spent synchronized pre- and post-Ironmen race. Outliers are presented by star (*).  The length of synchronizations, given by the length of the plateaus  is larger in the post-race indicating that the athlete is more relaxed after the competition and shows a better synchronization between cardiac and respiratory systems.   }
\label{pre_post}
\end{figure}

Figure ~\ref{pre_post} displays box plots for the cardiorespiratory synchronization times both before and after the Ironman race. The figure shows a clear difference  with the synchronization times being significantly higher post-race with a $p$-value of 0.009.
%%%%%%%%%%%%%%%%%%
%%%%%%%%%%%%%%%%%%%%%%%%%%%%%

These results led to the conclusion that the synchronization is  stronger  post-competition.
%%%%%%%%%%%%%%%%%%%%%%%%%%%
%{\color{red}
%\subsection{Stroop Test}

%%%%%%%%%%%%%%%%%%%%

%%%%%%%%%%%%%%%%%%%%%%%%%

%\subsection{Stroop Test results} 
%%%%%%%%%%%%%%%%%%%%%%%%%%%%%%%%%
%\textcolor{red}{The stroop test results for all the %participants can be filled here.}
%%%%%%%%%%%%%%%%%%%%%%

\section{Discussion and Conclusion}
\label{Conclusions}
%%%%%%%%%%%%%%%%%%%%%%%

A new method for visualising the synchronizations between the cardiorespiratory system was proposed through the implementation of  EMD and HHT. The moving variance also allows quantification of the stability of these synchronized regions.

Strong synchronizations were observed in the Ironman athletes post-competition, these periods were significantly longer and more pronounced than the synchronized regions witnessed prior to the competition for 13 out of the 14 athletes.  Although the Stroop test was impeding any conscious efforts to regulate the cardiorespiratory systems, unconsciously the body's need to recover homeostasis after the race meant that the control mechanisms are still working to regulate the heart and breathing rates - in order to restore them to a normal rate.

The Ironman competitors displayed the highest levels of synchronization during periods when their body's were recovering from a state of stress. This is  contrary to our hypothesis because  the athletes showed longer, more stable periods of synchronization, presumably partly due to a superior level of fitness and respiratory control.  

Another factor to consider is the amount of stress the individuals are recovering from, for example the effects of an Ironman competition are far greater than those of a single Stroop test.  Therefore the recovery phase after competition is much more important for restoring and maintaining homeostasis.  This heightened importance, we believe, is another reason for stronger synchronizations.

 Finally, seeing such high levels of synchronization in the Ironman athletes after competition - when completing a Stroop test - indicates the controlled breathing is not a requirement for cardiorespiratory synchronizations.   Moreover, the synchronizations seen suggest even more cardiorespiratory coordination in the absence of conscious control.
%%%%%%%%%%

\section {Acknowledgements} 
%%%%%%%%%%%%%%

MA, PH and LR thank the FP7 research  Project Models for Ageing  and Technological Solutions for Improving and Enhancing the Quality of Life (MATSIQEL),  under Grant FP7-PEOPLE-IRSES-247541 and the Medical Research Council of South Africa, the University of Cape Town Harry Crossley and Nellie Atkinson Staff Research Funds for the partial support of this work.
\bibliographystyle{apalike}  
\bibliography{sync_refs}

\end{document}